# A description of the pseudorapidity distributions in heavy ion collisions at RHIC and LHC energies

Z. J. Jiang[*], Y Zhang, H. L. Zhang and H. P. Deng

*College of Science, University of Shanghai for Science and Technology, Shanghai 200093, China*

The charged particles produced in nucleus-nucleus collisions are classified into two parts：One is from the hot and dense matter created in collisions. The other is from leading particles. The hot and dense matter is assumed to expand and generate particles according to BJP hydrodynamics, a theory put forward by A. Bialas, R. A. Janik and R. Peschanski. The leading particles are argued to possess a Gaussian rapidity distribution with the normalization constant equaling the number of participants. A comparison is made between the theoretical results and the experimental measurements performed by BRAHMS and PHOBOS Collaboration at BNL-RHIC in Au-Au and Cu-Cu collisions at $\sqrt{s_{NN}}$ =200 GeV and by ALICE Collaboration at CERN-LHC in Pb-Pb collisions at $\sqrt{s_{NN}}$ =2.76 TeV. The theoretical results are well consistent with experimental data.

## Ⅰ. Introduction

Relativistic hydrodynamics, since its success in describing the elliptic flow and multiplicity production in heavy ion collisions [1-4], has now been widely taken as one of the best tools for understanding the space-time evolution of the matter created in collisions [5-11].

Owing to the tremendous complexity of hydrodynamic equations, the most analytical investigations are, up till now, mainly limited to the hydrodynamics of 1+1 dimensions, which was first considered by Landau in the context of high energy collisions [12]. The first exact analytical solution of 1+1 hydrodynamics was given by I. M. Khalatnikov about 60 years ago [13] and was used later by L. D. Landau in his hydrodynamic model study [14]. This solution is for an accelerated system with fluid being assumed as a massless perfect one and initially at rest. The obtained result is a very unpleasant one since it is presented in a rather implicit way. However, the good news is that, from this complicated solution, Landau managed to extract approximately the rapidity distributions of the charged particles, which are in generally consistent with the observations made at BNL-RHIC [15-17].

---

[*] E-mail: jzj265@163.com

The second exact analytical solution of 1+1 dimensional hydrodynamics is given by R. C. Hwa about 40 years ago [18]. This solution is for an accelerationless system with Lorentz invariant initial condition. The result got in this way is simple and explicit. From this solution, J. D. Bjorken was able to get a simple estimate for the initial energy density achieved in collisions from the final observables [19]. This makes the energy density be measurable in experiment. However, the sad news is that the invariant rapidity distributions obtained from this model are at variance with experimental observations.

Among other theoretical models, it is worth to notice that a series of exact solutions of relativistic hydrodynamics has been found in recent years. By generalizing Hwa-Bjorken in-out ansatz for fluid trajectories, Ref. [20] presented a family of exact solutions. The typical characteristic of this model is that it interpolates between Hwa-Bjorken picture and Landau one. By taking into account the work done by the fluid elements on each other, Refs. [21-23] generalized the Hwa-Bjorken model for an accelerationless system to the one for an accelerated system, and obtained a new class of exact analytical solutions of relativistic hydrodynamics. What is more important is that, unlike Hwa-Bjorken model, the rapidity distributions derived from this model are well consistent with experimental data. By taking advantage of the scheme of Khalatnikov potential, Ref. [24] solved analytically the hydrodynamic equations and gave a pack of simple exact solutions for the perfect fluid fitting the linear equation of state. By using scalar field theory, Ref. [25] got a set of exact solutions describing non-stationary and inhomogeneous flows of the perfect fluid under different linear and non-linear equations of state. By applying the lattice QCD equation of state, Ref. [26] acquired a group of exact solutions for an accelerationless system with ellipsoidal symmetry.

A direct application of the 1+1 dimensional hydrodynamics is the analysis of the pseudorapidity distributions of the charged particles produced in heavy ion collisions. In this paper, we shall use the 1+1 dimensional hydrodynamics put forward by A. Bialas, R. A. Janik and R.

Peschanski in Ref [20] to discuss the pseudorapidity distributions of the charged particles produced in heavy ion collisions. For the convenience of following applications, we shall call this theory as BJP hydrodynamics. The appealing features of this model are that it unifies the characteristics of Hwa-Bjorken and Landau two famous hydrodynamic models. The solution of this model is exactly presented in an explicit and simple form, and the rapidity distribution of charged particles derived from this solution can be expressed in an analytical form with only two free parameters.

The key ingredients of this model are illustrated in section Ⅱ. In section Ⅲ, a comparison is made between the theoretical results and experiment measurements carried out by BRAHMS and PHOBOS Collaboration at BNL-RHIC in Au-Au and Cu-Cu collisions at $\sqrt{s_{NN}}$ =200 GeV [15-17, 27] and by ALICE Collaboration at CERN-LHC in Pb-Pb collisions at $\sqrt{s_{NN}}$ =2.76 TeV [28]. The comparison indicates that, besides particles coming from fluid evolution, leading particles are also essential in describing experimental data [29]. Only after both of them are taken into account together, can the experimental observations be described properly. The last section is traditionally about conclusions.

## Ⅱ. A brief introduction to BJP hydrodynamics

Here, for the purpose of completion and applications, we shall firstly give a brief introduction to the main points of BJP hydrodynamics.

(1) The hot and dense matter created in collisions is taken as a perfect fluid, which meets the equation of state

$$\varepsilon = gp, \tag{1}$$

where $\varepsilon$ is the energy density, $p$ is the pressure, and $1/\sqrt{g} = c_s$ is the speed of sound.

Eq. (1) makes the expansion equation of fluid along longitudinal direction (taken as $z$ axis) possess the form as

$$\frac{e^{2y}-1}{2}(g+1)\partial_+ p + e^{2y}(g+1)p\partial_+ y + \frac{1-e^{-2y}}{2}(g+1)\partial_- p +$$
$$e^{-2y}(g+1)p\partial_- y + \partial_+ p - \partial_- p = 0,$$

$$\frac{e^{2y}+1}{2}(g+1)\partial_+ p + e^{2y}(g+1)p\partial_+ y + \frac{1+e^{-2y}}{2}(g+1)\partial_- p -$$
$$e^{-2y}(g+1)p\partial_- y - \partial_+ p - \partial_- p = 0,$$
(2)

where $y$ is ordinary rapidity. $\partial_+$ and $\partial_-$ denote the compact notations of partial derivatives with respect to light-cone coordinates $z_\pm = t \pm z = \tau e^{\pm \eta_s}$, $\eta_s$ is the space-time rapidity, and $\tau = \sqrt{z_+ z_-}$ is the proper time.

(2) Eq. (2) is a complicated, non-linear and coupled one. In order to solve it, the relation between $y$ and $\eta_s$ is generalized to

$$2y = \ln u_+ - \ln u_- = \ln F_+(z_+) - \ln F_-(z_-), \qquad (3)$$

where $u_\pm = e^{\pm y}$ is the light-cone components of 4-volicity of fluid. $F_\pm(z_\pm)$ are a priori arbitrary function. In case of $F_\pm(z_\pm) = z_\pm$, Eq. (3) reduces to $y = \eta_s$, returning to the boost-invariant picture of Hwa-Bjorken. Otherwise, Eq. (3) describes the non-boost-invariant geometry of Landau. Accordingly, Eq. (3) unifies the Hwa-Bjorken and Landau hydrodynamics. It paves a way between these two models.

By using Eq. (3), Eq. (2) can be rewritten as

$$g\partial_+ \ln p = -\frac{(g+1)^2}{4}\frac{f'_+}{f_+} + \frac{g^2-1}{4}\frac{f'_-}{f_+},$$
$$g\partial_- \ln p = -\frac{(g+1)^2}{4}\frac{f'_-}{f_-} + \frac{g^2-1}{4}\frac{f'_+}{f_-},$$
(4)

where $f = F/H$, and $H$ is an arbitrary constant. The above equation is now solvable. Its solution is

$$s = s_0 \left(\frac{T}{T_0}\right)^g = s_0 \left(\frac{p}{p_0}\right)^{\frac{g}{g+1}}$$
$$= s_0 \exp\left[-\frac{g+1}{4}\ln(f_+ f_-) + \frac{g-1}{2}\sqrt{\ln f_+ \ln f_-}\right] = s_0 \exp(-g\theta), \qquad (5)$$
$$\theta = \ln\left(\frac{T_0}{T}\right) = \frac{g+1}{4g}\ln(f_+ f_-) - \frac{g-1}{2g}\sqrt{\ln f_+ \ln f_-},$$

where $s$ and $T$ are respectively the entropy density and temperature of fluid, $s_0$ and $T_0$ are the initial scales of them.

(3) The freeze-out of fluid takes place at a space-like hypersurface with a fixed temperature of $T_{FO}$. From this point together with the direct proportional relation between the number of charged particles and entropy, we can get the rapidity distribution of the charged particles

$$\frac{dN_{Fluid}(z_+, z_-)}{dy} = Ce^{-\frac{1}{4}(g-1)\left(\sqrt{\ln f_+} - \sqrt{\ln f_-}\right)^2} \frac{\sqrt{\ln f_+} + \sqrt{\ln f_-}}{(g+1)\sqrt{\ln f_+ \ln f_-} - \frac{g-1}{2}(\ln f_+ + \ln f_-)}\bigg|_{FO}, \quad (6)$$

where $C$, independent of $z_+$ and $z_-$, is an overall normalization constant.

(4) The right hand side of Eq. (6) is evaluated at a space-like hypersurface with a fixed temperature of $T_{FO}$. Known from Eq. (5), this means that, for charged particles produced in heavy ion collisions, $f_+$ and $f_-$ should meet relation

$$\theta_{FO} = \ln\left(\frac{T_0}{T_{FO}}\right) = \frac{g+1}{4g}\ln(f_+ f_-) - \frac{g-1}{2g}\sqrt{\ln f_+ \ln f_-}, \quad (7)$$
$$\ln f_+ - \ln f_- = 2y.$$

From this equation, we can get solution

$$\frac{g-1}{4}\left(\sqrt{\ln f_+} - \sqrt{\ln f_-}\right)^2 = \frac{g-1}{2}\left(\theta_{FO} - \sqrt{\theta_{FO}^2 - y^2/g}\right),$$
$$\sqrt{\ln f_+} + \sqrt{\ln f_-} = \sqrt{2}y\frac{1}{\left(\theta_{FO} - \sqrt{\theta_{FO}^2 - y^2/g}\right)^{\frac{1}{2}}}, \quad (8)$$
$$(g+1)\sqrt{\ln f_+ \ln f_-} - \frac{g-1}{2}(\ln f_+ + \ln f_-) = 2g\sqrt{\theta_{FO}^2 - y^2/g}.$$

Inserting them into Eq. (6), we finally obtain

$$\frac{dN_{Fluid}(\sqrt{s_{NN}}, y)}{dy} = C(\sqrt{s_{NN}})e^{-\frac{g-1}{2}\left(\theta_{FO} - \sqrt{\theta_{FO}^2 - y^2/g}\right)} \times \frac{y}{\left[\left(\theta_{FO} - \sqrt{\theta_{FO}^2 - y^2/g}\right)\left(\theta_{FO}^2 - y^2/g\right)\right]^{\frac{1}{2}}}. \quad (9)$$

It is the rapidity distribution of the charged particles resulting from the freeze-out of fluid. It is an analytical function of $y$ with two free parameters $\theta_{FO}$ and $g$.

**III. Comparison with experimental measurements and the rapidity distributions**

**of leading particles**

Here, we shall use Eq. (9) in heavy ion collisions. It is evident that the parameter $\theta_{FO}$ in Eq. (9) should be the function of incident energy and centrality cut. Its specific value can be fixed by comparing with experimental data. The speed of sound in Eq. (9) takes the value of $c_s = 1/\sqrt{g} = 0.35$ [30-33], which is almost independent of energy and collision system.

Figure 1 shows the rapidity distributions for $\pi^+$, $\pi^-$, $K^+$, $K^-$, $p$ and $\bar{p}$ produced in central Au-Au collisions at $\sqrt{s_{NN}}$ =200 GeV. The scattered symbols are the experimental data [15-17]. The solid curves are the theoretical results from Eq. (9). In calculations, the parameter $\theta_{FO}$ takes the value of $\theta_{FO}$ =2.23. It can be seen from this figure that, except for proton $p$, Eq. (9) fits well with experimental measurements. For proton $p$, experimental data show an evident uplift in the rapidity interval between $y = 2.0$ and 3.0. This may be resulting from parts of leading particles, which is out of the scope of hydrodynamics. Hence, in order to describe experiments properly, we should take these leading particles into account separately.

Considering that, for a given incident energy, different leading particles resulting from each nucleus-nucleus collision have on the average the same amount of energy, then, according to the central limit theorem [34, 35], the leading particles should follow the Gaussian rapidity distribution. That is

$$\frac{dN_{Lead}(b,\sqrt{s_{NN}},y)}{dy} = \frac{N_{Lead}(b,\sqrt{s_{NN}})}{\sqrt{2\pi}\sigma} \exp\left\{-\frac{\left[|y|-y_0(b,\sqrt{s_{NN}})\right]^2}{2\sigma^2}\right\}, \quad (10)$$

where $y_0(b,\sqrt{s_{NN}})$ and $\sigma$ are respectively the central position and width of distribution. In fact, as is known to all, the rapidity distributions of any charged particles produced in heavy ion collisions can be well represented by Gaussian form [15-17; also confer the shapes of the curves in Figure 1].

It is evident that $y_0(b,\sqrt{s_{NN}})$ should increase with centrality cuts and incident energies. While, $\sigma$ should not apparently depend on them. This is due to the fact that $\sigma$ is determined by the relative energy differences among leading particles, and such energy differences should not be too much for different centrality cuts and energies. $N_{\text{Lead}}\left(b,\sqrt{s_{\text{NN}}}\right)$ in Eq. (10) is the number of leading particles.

It is well known that, in nucleon-nucleon, such as *p-p* collisions, there are two leading particles. One is in projectile fragmentation region, and the other is in target fragmentation region. Then, in nucleus-nucleus collisions, the leading particles should be those nucleons which participate in collisions, the so-called participants, which locate separately at projectile and target fragmentation regions. For collisions between two identical nuclei, each nucleus should have about the same number of participants. Hence, the number of leading particles appearing in projectile or target fragmentation region should be

$$N_{\text{Lead}}\left(b,\sqrt{s_{\text{NN}}}\right)=\frac{N_{\text{Part}}\left(b,\sqrt{s_{\text{NN}}}\right)}{2},$$

where $N_{\text{Part}}\left(b,\sqrt{s_{\text{NN}}}\right)$ is the number of total participants in two nuclei, which can be evaluated by formula [36]

$$N_{\text{Part}}\left(b,\sqrt{s_{\text{NN}}}\right)=\int n_{\text{Part}}\left(b,\sqrt{s_{\text{NN}}},s\right)\mathrm{d}^2 s, \quad (11)$$

where $s$ is the coordinates in the overlap region measured from the center of the projectile nucleus.

$$n_{\text{Part}}\left(b,\sqrt{s_{\text{NN}}},s\right)$$
$$=T_A(s)\left\{1-\exp\left[-\sigma_{\text{NN}}^{\text{in}}\left(\sqrt{s_{\text{NN}}}\right)T_B(s-b)\right]\right\}+T_B(s-b)\left\{1-\exp\left[-\sigma_{\text{NN}}^{\text{in}}\left(\sqrt{s_{\text{NN}}}\right)T_A(s)\right]\right\},$$

where $\sigma_{\text{NN}}^{\text{in}}\left(\sqrt{s_{\text{NN}}}\right)$ is the inelastic nucleon-nucleon cross section. It increases slowly with energies. Such as, for $\sqrt{s_{\text{NN}}}$ =200 GeV, $\sigma_{\text{NN}}^{\text{in}}$ =42 mb [37], and for $\sqrt{s_{\text{NN}}}$ =2.76 TeV, $\sigma_{\text{NN}}^{\text{in}}=64\pm 5$ mb [38]. The subscripts A and B in above equation denote the projectile and target nucleus, respectively. $T(s)$ is the thickness function defined as

$$T(s)=\int \rho(s,z)\mathrm{d}z, \quad (12)$$

where

$$\rho(r)=\frac{\rho_0}{1+\exp[(r-r_0)/a]}$$

is the Woods-Saxon distribution of nuclear density. $a$ and $r_0$ are respectively the skin depth and radius of nucleus. In this paper, they take the values of $a$=0.54 fm and $r_0 = 1.12 A^{1/3} - 0.86 A^{-1/3}$ fm [39], where $A$ is the mass number of nucleus.

Tables I and II show the mean numbers of total participants in different centrality Au-Au and Cu-Cu collisions at $\sqrt{s_{NN}}$ = 200 GeV and Pb-Pb collisions at $\sqrt{s_{NN}}$ =2.76 TeV. The numbers with and without errors are those given by experiments [27, 28] and Eq. (11), respectively. Due to the space constraints, Table I only shows the numbers in the first nine centrality cuts. It can be seen that both sets of numbers coincide well.

Having the rapidity distributions of Eqs. (9) and (10), the pseudorapidity distribution measured in experiments can be expressed as [40]

$$\frac{dN(b,\sqrt{s_{NN}},\eta)}{d\eta} = \sqrt{1 - \frac{m^2}{m_T^2 \cosh^2 y}} \frac{dN(b,\sqrt{s_{NN}},y)}{dy}, \qquad (13)$$

$$y = \frac{1}{2}\ln\left[\frac{\sqrt{p_T^2 \cosh^2 \eta + m^2} + p_T \sinh\eta}{\sqrt{p_T^2 \cosh^2 \eta + m^2} - p_T \sinh\eta}\right], \qquad (14)$$

where $p_T$ is the transverse momentum, $m_T = \sqrt{m^2 + p_T^2}$ is the transverse mass, and

$$\frac{dN(b,\sqrt{s_{NN}},y)}{dy} = \frac{dN_{Fluid}(b,\sqrt{s_{NN}},y)}{dy} + \frac{dN_{Lead}(b,\sqrt{s_{NN}},y)}{dy} \qquad (15)$$

is the total rapidity distribution from both fluid evolution and leading particles.

Substituting Eq. (15) into (13), we can get the pseudorapidity distributions of the charged particles. Figures 2, 3 and 4 show such distributions in different centrality Au-Au and Cu-Cu collisions at $\sqrt{s_{NN}}$ =200 GeV and Pb-Pb collisions at $\sqrt{s_{NN}}$ =2.76 TeV, respectively. The solid dots in figures are the experimental measurements [27, 28]. The dashed curves are the results got from BJP hydrodynamics of Eq. (9). The dotted curves are the results obtained from leading particles of Eq. (10). The solid curves are the results achieved from Eq. (15), that is, the sums of dashed and dotted curves. It can be seen that the theoretical results are well consistent with experimental measurements.

Experiments have shown that the overwhelming majority of charged particles produced in heavy ion collisions at high energy consists of pions, kaons and protons with proportions of about 83%, 12% and 5%, respectively [41]. These proportions are not evidently dependent on colliding energies and systems. Furthermore, for a given incident energy, the transverse momentum $p_T$ changes very slowly with centrality cuts. For a specific type of charged particle, it can be well taken to be a constant for centrality cuts from 0%–55%, which we are interested in. In Au-Au collisions at $\sqrt{s_{NN}}$ =200 GeV, this constant is about 0.45, 0.65, and 0.93 GeV/$c$ for pions, kaons, and protons, respectively. In calculations, the mass and transverse momentum in Eqs. (13) and (14) take the values of $m=0.22$ GeV and $p_T=0.50$ GeV/$c$ in Au-Au and Cu-Cu collisions at $\sqrt{s_{NN}}$ =200 GeV, which are the mean values of those of pions, kaons and protons. In Pb-Pb collisions at $\sqrt{s_{NN}}$ =2.76 TeV, they take the values of $m=0.22$ GeV and $p_T=0.62$ GeV/$c$. Here, $p_T$ takes the mean value from experimental measurements [42, 43].

The parameter $\theta_{FO}$ in Eq. (9) takes the value of 2.80 in the first three centrality cuts, 2.98 in the following six ones and 3.17 in the last two ones in Au-Au collisions. In Cu-Cu collisions, $\theta_{FO}$ takes the value of 2.95 in the first three centrality cuts, 3.15 in the following six ones and 3.53 in the last three ones. In Pb-Pb collisions, $\theta_{FO}$ takes the value of 5.95 for all four centrality cuts. It can be seen that $\theta_{FO}$ increases with centrality cuts and incident energies. While, for a given centrality cut and energy, $\theta_{FO}$ decreases with increasing nucleus size. The width parameter $\sigma$ in Eq. (10) takes the values of 0.85 in different centrality Au-Au, Cu-Cu and Pb-Pb collisions. It is independent of centrality cut, energy and collision system. The center parameter $y_0$ in Eq. (10) takes the values as listed in Tables Ⅰ and Ⅱ. As mentioned above, $y_0$ increases with centrality cuts and energies. While, as it can be seen from Table Ⅰ that, for a given centrality cut and incident energy, $y_0(b,\sqrt{s_{NN}})$ decreases with increasing nucleus size. This can be understood if we notice the fact that the larger the nucleus size, the more collisions will the participants undergo. Hence, the final leading particles will lose more energy or have smaller $y_0(b,\sqrt{s_{NN}})$. The fitting value of $y_0=2.63$ in the top 3% most central Au-Au collisions is in accordance with the experimental observation shown in Fig. 1, which indicates that the leading particles are mainly in the range between $y = 2$ and 3. Experimental investigations also have shown that [44], in the top 5% most central Au-Au collisions at $\sqrt{s_{NN}} = 200$ GeV, the rapidity loss of participants is up to $<\delta y> \approx 2.45$, then the leading particles should locate at

$$y_0 = y_{beam} - <\delta y> = 5.36 - 2.45 = 2.91.$$

Seeing that the smaller centrality cut considered in our analysis, our above fitting result is also consistent with this measurement.

## Ⅳ. Conclusions

The charged particles in nucleus-nucleus collisions consist of two parts: One is from the freeze-out of the hot and dense matter created in collisions. The other is from the leading particles which are less understood.

The hot and dense matter is supposed to expand according to BJP hydrodynamics which provides us a set of exact solutions for a perfect fluid with linear equation of state. These solutions are essential in deriving the analytical rapidity distributions of the charged particles produced by the freeze-out of fluid at the space-like hypersurface with a fixed temperature of $T_{FO}$. The rapidity distributions got in this way have two parameters $g = 1/c_s^2$ and $\theta_{FO} = \ln(T_0/T_{FO})$ with $g$ taken the value from experimental measurements and $\theta_{FO}$ determined by comparing the theoretical predictions with experimental data.

For leading particles, we assume that the rapidity distribution of them possesses a Gaussian form with the normalization constant being equal to the number of participants, which can be figured out in theory. This assumption is based on the consideration that, for a given incident energy, the leading particles have about the same energy, and coincides with the fact that any kind of the charged particles produced in collisions takes on well the Gaussian form of rapidity distribution. It is interested to notice that the width $\sigma$ of Gaussian rapidity distribution is irrelevant to centrality cuts, energies and collision systems.

Comparing with experimental measurements carried out by BRAHMS and PHOBOS Collaboration at BNL-RHIC in Au-Au and Cu-Cu collisions at $\sqrt{s_{NN}} = 200$ GeV and by ALICE Collaboration at CERN-LHC in Pb-Pb collisions at $\sqrt{s_{NN}} = 2.76$ TeV, we can see that the total contributions from both BJP hydrodynamics and leading particles are well consistent with experimental data.

## Acknowledgments

This work is supported by the Transformation Project of Science and Technology of Shanghai Baoshan District with Grant No. CXY-2012-25.

**Table and figure captions**

**Table I**

The mean numbers of total participants $\bar{N}_{Part}$ and the central positions $y_0$ of Gaussian rapidity distributions in different centrality Au-Au and Cu-Cu collisions at $\sqrt{s_{NN}} = 200$ GeV. The numbers with and without errors are respectively the results given by PHOBOS Collaboration at BNL-RHIC [27] and Eq. (11).

**Table II**

The mean numbers of total participants $\bar{N}_{Part}$ and the central positions $y_0$ of Gaussian rapidity distributions in different centrality Pb-Pb collisions at $\sqrt{s_{NN}} = 2.76$ TeV. The numbers with and without errors are respectively the results given by ALICE Collaboration at CERN-LHC [28] and Eq. (11).

**Figure 1**

The rapidity distributions of specified charged particles in central Au-Au collisions at $\sqrt{s_{NN}} = 200$ GeV. The scattered symbols are the experimental measurements [15-17]. The solid curves are the results from BJP hydrodynamics of Eq. (9).

**Figure 2**

The pseudorapidity distributions of the charged particles produced in different centrality Au-Au collisions at $\sqrt{s_{NN}} = 200$ GeV. The solid dots are the experimental measurements [27]. The dashed curves are the results from BJP hydrodynamics of Eq. (9). The dotted curves are the results from leading particles of Eq. (10). The solid curves are the sums of dashed and dotted ones.

**Figure 3**

The pseudorapidity distributions of the charged particles produced in different centrality Cu-Cu collisions at $\sqrt{s_{NN}} = 200$ GeV. The solid dots are the experimental measurements [27]. The dashed curves are the results from BJP hydrodynamics of Eq. (9). The dotted curves are the results from leading particles of Eq. (10). The solid curves are the sums of dashed and dotted ones.

**Figure 4**

The pseudorapidity distributions of the charged particles produced in different centrality Pb-Pb collisions at $\sqrt{s_{NN}} = 2.76\,\text{TeV}$. The solid dots are the experimental measurements [28]. The dashed curves are the results from BJP hydrodynamics of Eq. (9). The dotted curves are the results from leading particles of Eq. (10). The solid curves are the sums of dashed and dotted ones.

**Table I**

| Centrality Cut (%) | 0-3 | 3-6 | 6-10 | 10-15 | 15-20 | 20-25 | 25-30 | 30-35 | 35-40 |
|---|---|---|---|---|---|---|---|---|---|
| $\bar{N}_{\text{Part}}(\text{Au-Au})$ | 359.44 | 324.50 | 288.74 | 248.96 | 210.98 | 178.24 | 149.78 | 124.92 | 103.22 |
| | 361±11 | 331±10 | 297±9 | 255±8 | 215±7 | 180±7 | 150±6 | 124±6 | 101±6 |
| $\bar{N}_{\text{Part}}(\text{Cu-Cu})$ | 109.92 | 99.76 | 89.00 | 76.70 | 64.74 | 54.40 | 45.40 | 37.62 | 30.88 |
| | 108±4 | 101±3 | 91±3 | 79±3 | 67±3 | 57±3 | 48±3 | 40±3 | 33±3 |
| $y_0(\text{Au-Au})$ | 2.63 | 2.67 | 2.70 | 2.72 | 2.78 | 2.81 | 2.96 | 2.97 | 3.05 |
| $y_0(\text{Cu-Cu})$ | 2.75 | 2.78 | 2.80 | 2.93 | 2.94 | 2.95 | 2.96 | 2.97 | 3.05 |

**Table II**

| Centrality Cut (%) | 0-5 | 5-10 | 10-20 | 20-30 |
|---|---|---|---|---|
| $\bar{N}_{\text{Part}}(\text{Pb-Pb})$ | 381.56 | 327.70 | 261.90 | 189.78 |
| | 383±3 | 330±5 | 261±4 | 186±4 |
| $y_0(\text{Pb-Pb})$ | 3.38 | 3.41 | 3.44 | 3.48 |

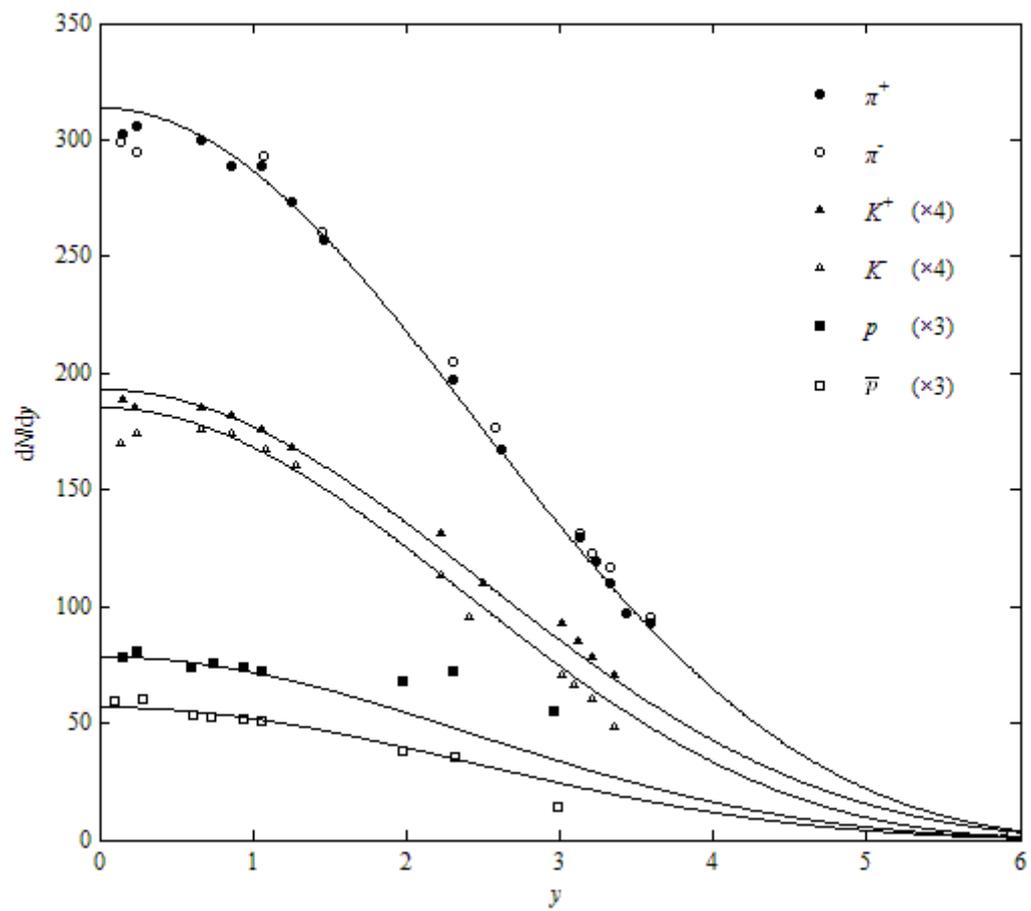

**Figure 1**

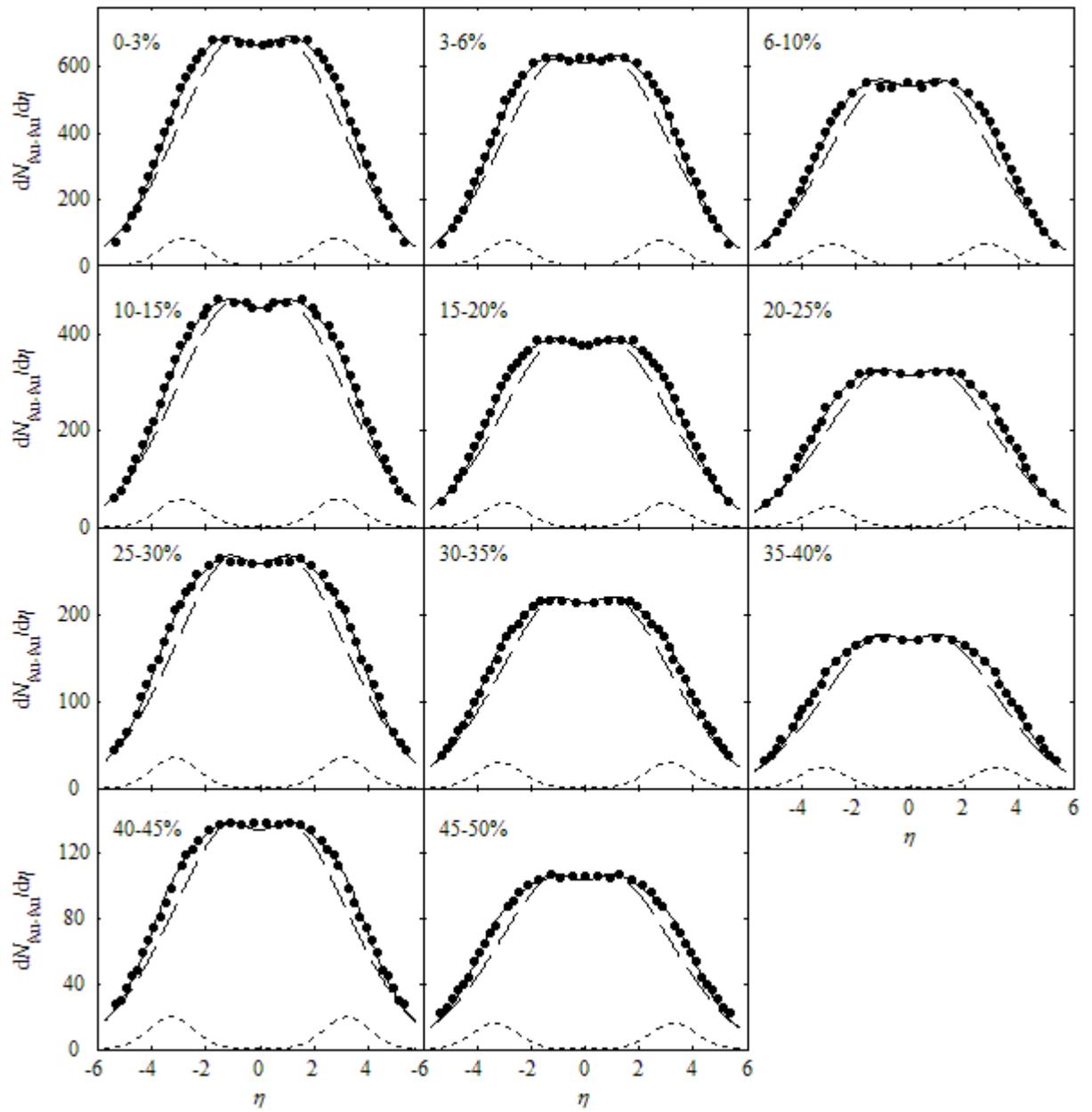

**Figure 2**

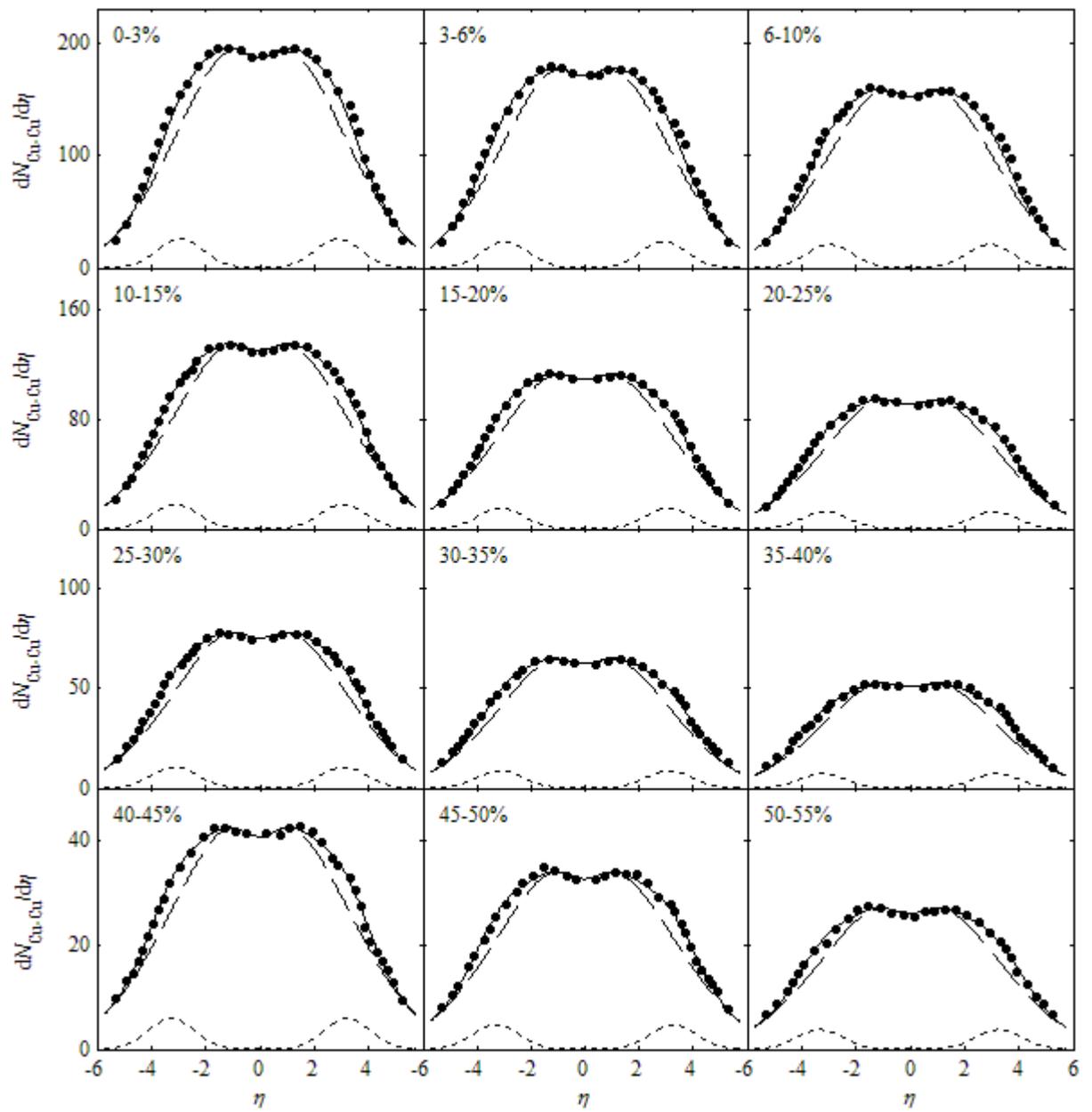

**Figure 3**

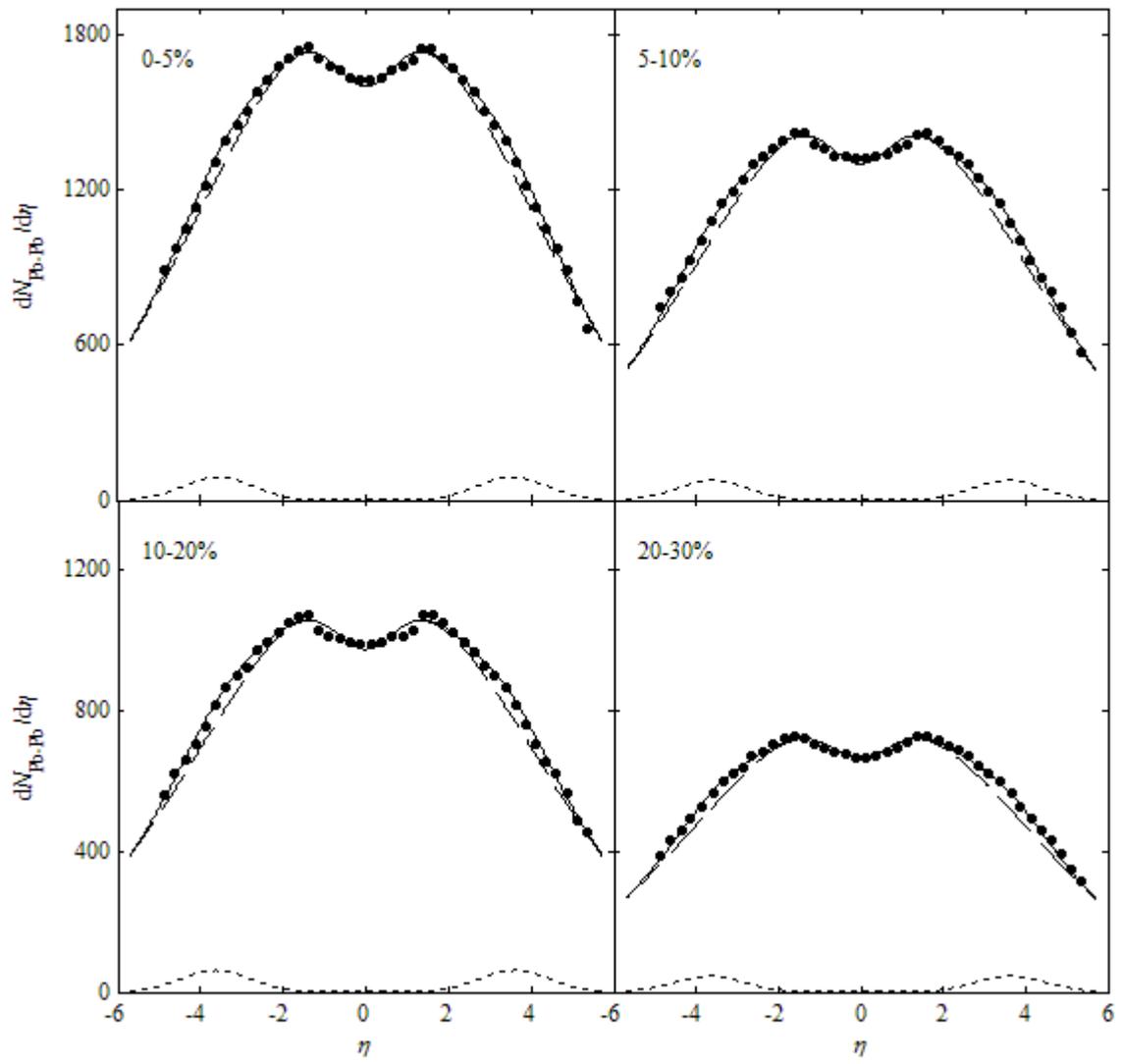

**Figure 4**